# Stabilizing the forming process in unipolar resistance switching using an improved compliance current limiter

S. B. Lee,[1] S. H. Chang,[1] H. K. Yoo,[1] and B. S. Kang[2,a]

[1]*ReCFI, Department of Physics and Astronomy, Seoul National University, Seoul 151-747, Republic of Korea*

[2]*Department of Applied Physics, Hanyang University, Ansan, Gyeonggi-do 426-791, Republic of Korea*

---

[a]Electronic mail: bosookang@hanyang.ac.kr

The high reset current $I_R$ in unipolar resistance switching now poses major obstacles to practical applications in memory devices. In particular, the first $I_R$-value after the forming process is so high that the capacitors sometimes do not exhibit reliable unipolar resistance switching. We found that the compliance current $I_{comp}$ is a critical parameter for reducing $I_R$-values. We therefore introduced an improved, simple, easy to use $I_{comp}$-limiter that stabilizes the forming process by drastically decreasing current overflow, in order to precisely control the $I_{comp}$- and subsequent $I_R$-values.

**I. INTRODUCTION**

Unipolar resistance switching (RS) shows reversible bistable resistance states that depend on the magnitude of the same-polarity bias voltage in thin film capacitors. It has attracted renewed interest due to its potential application in nonvolatile resistance random access memory (RRAM) devices.[1–4] However, to use unipolar RS as a commercial nonvolatile memory device, several technical issues need to be resolved.

The major challenge is reducing the high current required by the reset process to change from the low (LRS) to the high resistance state (HRS).[5–7] The reset current $I_R$ is usually so high that unipolar RRAM requires a very high density of oxide diodes switch elements to produce a reliable three-dimensional array structure.[8] Therefore, high $I_R$ is not merely a power consumption issue, but also a serious constraint on the feasibility of practical device operation.

The high $I_R$ is closely related to the breakdown-like failure in unipolar RS capacitors. Even if the capacitors are fabricated on the same wafer, they show a large $I_R$ distribution after the forming process, which is the first dielectric breakdown-like resistance change from pristine state to LRS. At the forming process, conducting filaments originally have percolating connectivity inside the film.[2,4,7] The subsequent unipolar RS is widely accepted to occur due to the formation and rupture of these conducting filaments inside the film.[1–4] After the forming process, some of the capacitors show an $I_R$ high enough to exceed the capacity of typical measuring equipment. The unipolar RS in those capacitors was therefore regarded as permanently failed. Little attention thus far has been paid to this issue in the literature;[6] it has been attributed to sample quality without any evidence. Therefore, to consistently achieve reliable unipolar RS, the forming process should be studied systematically to obtain consistent control of the process.

We observed that a current overflow at the forming process plays an important role in the subsequent $I_R$-values and hence in the prevention of permanent breakdown after the forming process. In the following, we discuss our simple, easy to use method of overcoming current overflow.

## II. EXPERIMENTS

We fabricated polycrystalline Pt/NiO$_w$/Pt, Pt/SrTiO$_x$/Pt, Ti/SrTiO$_x$/Pt, Pt/TiO$_y$/Pt, and Pt/FeO$_z$/Pt capacitors, which showed unipolar RS. Details on the fabrication methods are well described elsewhere.[7,9,10] Current–voltage (*I*–*V*) characteristics were measured with a simple two-probe method using an Agilent 4155C semiconductor parameter analyzer (SPA). The bottom electrodes were grounded and the voltage was swept on the top electrodes for all the electrical measurements.

## III. RESULTS AND DISCUSSION

Figure 1 shows typical $I$–$V$ curves for Pt/NiO$_w$/Pt unipolar RS capacitors, which are highly insulating in the pristine state. When we applied a voltage of approximately 5–7 V, the current increased suddenly, electroforming the capacitors, as indicated by the blue line. Immediately after this forming process, the capacitor entered into a LRS. When the bias voltage increased above a reset voltage, the capacitor switched from the LRS to a HRS. As we further increased the bias voltage in the HRS, it reached a set voltage, and the film returned to the LRS at a still higher voltage, which is called a set process. For both the forming and set processes, we used a compliance current $I_{comp}$ to prevent permanent breakdown of the capacitors.

For the first reset process (red line) after the forming process, the $I_R$-value is around 50 mA despite the small $I_{comp}$ of 5 mA. In this measurement, we used an $I_{comp}$-limiter function provided by the SPA. Subsequently, even though the same $I_{comp}$ of 5 mA was used for the set processes, the $I_R$-values fluctuated between 5 mA and 20 mA, smaller than the first $I_R$-value but still larger than $I_{comp}$. After the forming process in some capacitors, the first $I_R$-value was larger than 100 mA, after which we failed to accomplish the reset process within the available $I_{comp}$-range of the SPA (≤ 100 mA). Many researchers have regarded those capacitors as undergoing a permanent breakdown. Therefore, the forming process and the subsequent reset process are important steps to obtain reliable unipolar RS. In the same measurement system, we also observed the problematically high $I_R$-value phenomenon after the forming process for many Pt/SrTiO$_x$/Pt, Ti/SrTiO$_x$/Pt, Pt/TiO$_y$/Pt, and Pt/FeO$_z$/Pt capacitors.

Recently, after performing conductive atomic-force microscopy studies on polycrystalline TiO$_y$ thin film surfaces, we reported that conducting filaments are locally formed during the forming and set processes.[2] At those moments, the total cross section of the conducting filaments is determined by the $I_{comp}$-values, so $I_R$-values should theoretically be proportional to $I_{comp}$-values.[7] However, as shown in Fig. 1, conventional $I_{comp}$-limiters such as the SPA of Fig. 2(a) could not block an abrupt change in current exactly at the forming and set processes.

How can the $I_{comp}$-values be precisely controlled to obtain the correct $I_R$-values? Many reports have addressed the discrepancies between the $I_{comp}$-values of the set processes and the $I_R$-values that occur as due to the parasitic capacitance between the unipolar RS capacitor and the $I_{comp}$-limiter.[6,11] Recently, Song *et al*. pointed out that this problem was due to the dissipation of capacitive charges that accumulated in the unipolar RS capacitor itself.[12] These two effects can also work together to induce these discrepancies.

To solve the discrepancies between the set processes $I_{comp}$-values and the $I_R$-values, Kinoshita *et al*. suggested that the cable connection between the capacitor and the $I_{comp}$-limiter should be as short as possible.[6] They also fabricated unipolar RS capacitors and $I_{comp}$-limiters on the same wafer. Although this latter technique should expect to be effective in solving the permanent breakdown problem after the forming process, it is too difficult to apply at a laboratory level. Therefore, we used a switching transistor (ST, 2N2369), connected externally as shown in Fig. 2(b), as the $I_{comp}$-limiter. An ideal ST rising time would be 6 ns, which is much shorter than the reported switching times of 10–200 ns in the set processes.[7,13] We kept the cable connection between the bottom electrode and the ST as short as possible.

As shown in Fig. 2(c), the first $I_R$-value after the forming process in the Pt/NiO$_w$/Pt capacitor is drastically reduced by using the ST $I_{comp}$-limiter. For $I_{comp}$ = 0.2 mA, the SPA (red line) measured an $I_R$-value of around 10 mA. The ST $I_R$-value (blue line), however, is approximately 0.5 mA at $I_{comp} \approx 0.2$ mA. After using the ST $I_{comp}$-limiter, all the capacitors successfully underwent reset processes after the forming process and showed reliable unipolar RS. During successive RS, $I_R$-values become stable at around 0.5 mA. Reduction of the first $I_R$-values by using a ST $I_{comp}$-limiter was also observed in Pt/SrTiO$_x$/Pt, Ti/SrTiO$_x$/Pt, Pt/TiO$_y$/Pt, and Pt/FeO$_z$/Pt capacitors.

To address the physical effect of the ST $I_{comp}$-limiter, we monitored the current-flow time evolution through the Pt/NiO$_w$/Pt capacitor at the forming process. Using a YOKOGAWA DL1740 digital oscilloscope, we measured the voltage change across a serially connected 50 Ω load resistor on the unipolar RS capacitor to calculate the current-flow.

Current overflow at the forming process decreases drastically by using a ST $I_{comp}$-limiter instead of a SPA; see Fig. 3. The maximum ST current-flow is around 1.5 mA at $I_{comp} \approx 0.2$ mA. The subsequent $I_R$-value was approximately 0.5 mA, as mentioned in Fig. 2(c). In contrast, for the SPA, the current-flow was above 14 mA for $I_{comp} = 0.2$ mA and the subsequent $I_R$-value was around 10 mA. The overflow time for the ST ($\approx 110$ ns) is much smaller than that for the SPA ($\approx 4.5$ $\mu$s). The large differences in current overflow for the ST versus the SPA causes the remarkable difference in the subsequent $I_R$-values; see Fig. 2(c).

Note that as shown in the inset of Fig. 3, even when we used the SPA $I_{comp}$-limiter,

the current overflow decreased drastically for the set process as compared to the forming process. As an example, in a set process with $I_{comp}$ = 0.2 mA, the maximum current flow and time were around 3 mA and 360 ns, respectively. Therefore, the $I_R$-values after the second reset process could become smaller than those of the first reset process; see Fig. 1. Understanding how the current overflows decrease at the set process is important. One possibility is that the HRS capacitive charges accumulated less than those in the pristine state. For the set processes, the reconnection of conducting filaments occurs at the sub-nanometer local regions.[2,4] The capacitive regions, which can accumulate capacitive charges, might be much smaller than those of the forming process.[12] Further study should help in elucidating this decrease in current overflow for the set process.

As shown in Fig. 4, the reduction of $I_R$-values in Pt/NiO$_w$/Pt capacitors when using the ST $I_{comp}$-limiter is also applicable to set processes. The relationship $I_R \propto I_{comp}$ holds $I_{comp}$ down to approximately 0.7 mA for the ST (blue squares), which is lower than that for the SPA (red triangles) by an order of magnitude. However, below $I_{comp} \approx 0.7$ mA, the ST $I_R$-values were saturated due to the current overflow, which is beyond the blocking ability of the ST. This relationship was also observed in Pt/SrTiO$_x$/Pt, Ti/SrTiO$_x$/Pt, Pt/TiO$_y$/Pt, and Pt/FeO$_z$/Pt capacitors, although the $I_R$-values were saturated at somewhat different values for the different film materials.

**IV. CONCLUSIONS**

We found that permanent breakdown after the forming process in unipolar resistance switching was due to current overflow, which could be attributed to the measurement system in a significant part. We solved this problem by using a switching transistor as an $I_{comp}$-limiter for both the forming and set processes. This improved $I_{comp}$-limiter decreased the minimum possible $I_R$-values by an order of magnitude compared to those when a conventional $I_{comp}$-limiter was used. This capability improvement in $I_{comp}$-limiting was attributable to the drastic decrease of current overflow in the forming and set processes. Our method is so simple and easy to use that it can be applied at the laboratory level.

**ACKNOWLEDGMENTS**

We acknowledge both the valuable discussions with and the fabrication of samples by Dr. M.-J. Lee and Dr. C. J. Kim at the Samsung Advanced Institute of Technology.

This research was supported by National Research Foundation of Korea (NRF) grants funded by the Korean government (MEST) (No. 2009-0080567 and No. 2010-0020416). B.S.K. was supported by the research fund of Hanyang University (HY-2009-N). S.B.L. acknowledges support from the Seoam Fellowship.

[1]R. Waser, R. Dittmann, G. Staikov, and K. Szot, Adv. Mater. (Weinheim, Ger.) **21**, 2632 (2009).

[2]S. C. Chae, J. S. Lee, S. Kim, S. B. Lee, S. H. Chang, C. Liu, B. Kahng, H. Shin, D.-W. Kim, C. U. Jung, S. Seo, M.-J. Lee, and T. W. Noh, Adv. Mater. (Weinheim, Ger.) **20**, 1154 (2008).

[3]S. H. Chang, J. S. Lee, S. C. Chae, S. B. Lee, C. Liu, B. Kahng, D.-W. Kim, and T. W. Noh, Phys. Rev. Lett. **102**, 026801 (2009).

[4]D.-H. Kwon, K. M. Kim, J. H. Jang, J. M. Jeon, M. H. Lee, G. H. Kim, X.-S. Li, G.-S. Park, B. Lee, S. Han, M. Kim, and C. S. Hwang, Nature Nanotechnol. **5**, 148 (2010).

[5]S.-E. Ahn, M.-J. Lee, Y. Park, B. S. Kang, C. B. Lee, K. H. Kim, S. Seo, D.-S. Suh, D.-C. Kim, J. Hur, W. Xianyu, G. Stefanovich, H. Yin, I.-K. Yoo, J.-H. Lee, J.-B. Park, I.-G. Baek, and B. H. Park, Adv. Mater. (Weinheim, Ger.) **20**, 924 (2008).

[6]K. Kinoshita, K. Tsunoda, Y. Sato, H. Noshiro, S. Yagaki, M. Aoki, and Y. Sugiyama, Appl. Phys. Lett. **93**, 033506 (2008).

[7]S. B. Lee, A. Kim, J. S. Lee, S. H. Chang, H. K. Yoo, T. W. Noh, B. Kahng, M.-J. Lee, C. J. Kim, and B. S. Kang, arXiv:1003.1390.

[8]B. S. Kang, S.-E. Ahn, M.-J. Lee, G. Stefanovich, K. H. Kim, W. X. Xianyu, C. B. Lee, Y. Park, I. G. Baek, and B. H. Park, Adv. Mater. (Weinheim, Ger.) **20**, 3066 (2008).

[9]S. Seo, M. J. Lee, D. H. Seo, E. J. Jeoung, D.-S. Suh, Y. S. Joung, I. K. Yoo, I. R. Hwang, S. H. Kim, I. S. Byun, J.-S. Kim, J. S. Choi, and B. H. Park, Appl. Phys. Lett. **85**, 5655 (2004).

[10]S. B. Lee, S. C. Chae, S. H. Chang, C. Liu, C. U. Jung, S. Seo, and D.-W. Kim, J. Korean Phys. Soc. **51**, S96 (2007).

[11]H. J. Wan, P. Zhou, L. Ye, Y. Y. Lin, T. A. Tang, H. M. Wu, and M. H. Chi, IEEE Electron Device Lett. **31**, 246 (2010).

[12]S. J. Song, K. M. Kim, G. H. Kim, M. H. Lee, J. Y. Seok, R. Jung, and C. S. Hwang, Appl. Phys. Lett. **96**, 112904 (2010).

[13]B. J. Choi, S. Choi, K. M. Kim, Y. C. Shin, C. S. Hwang, S.-Y. Hwang, S.-S. Cho, S. Park, and S.-K. Hong, Appl. Phys. Lett. **89**, 012906 (2006).

FIG. 1. (Color online) Current–voltage ($I$–$V$) curves in a Pt/NiO$_w$/Pt capacitor showing unipolar resistance switching. We used a semiconductor parameter analyzer (SPA) as a compliance current ($I_{comp}$) limiter for both the forming and set processes. Even with a small $I_{comp}$ = 5 mA at the forming process (blue line), the subsequent reset current $I_R$ was around 50 mA (red line). Subsequent $I_R$-values fluctuated greatly between 5 mA and 20 mA despite the small $I_{comp}$ = 5 mA value at the set processes.

FIG. 2. (Color online) Schematic diagrams of the electrical measurement systems. We used (a) a SPA or (b) a switching transistor (ST) as the $I_{comp}$-limiter. (c) Comparison of the first $I_R$-values after the forming process using SPA or ST $I_{comp}$-limiters. The ST $I_R$-value (blue line) becomes much smaller than that for the SPA (red line), even if we use approximately the same 0.2 mA $I_{comp}$.

FIG. 3. (Color online) Time evolutions of current-flow after the forming process. Even if we block the current-flow with an $I_{comp}$-limiter, the forming process shows a current overflow much larger than $I_{comp}$ = 0.2 mA. However, the ST current overflow (blue line) decreases drastically compared to that of the SPA (red line). The inset shows the current-flow time evolution for the set process with the SPA $I_{comp}$-limiter. Even with the SPA $I_{comp}$-limiter, the set process current overflow decreases more than that of the forming process.

FIG. 4. (Color online) $I_R \propto I_{comp}$ relationship for the set processes in unipolar resistance switching. This relationship is valid only above approximately 7 mA for the SPA $I_{comp}$-limiter (red triangles). However, the ST $I_R$ (blue squares) are saturated at a smaller value than for the SPA by an order of magnitude.

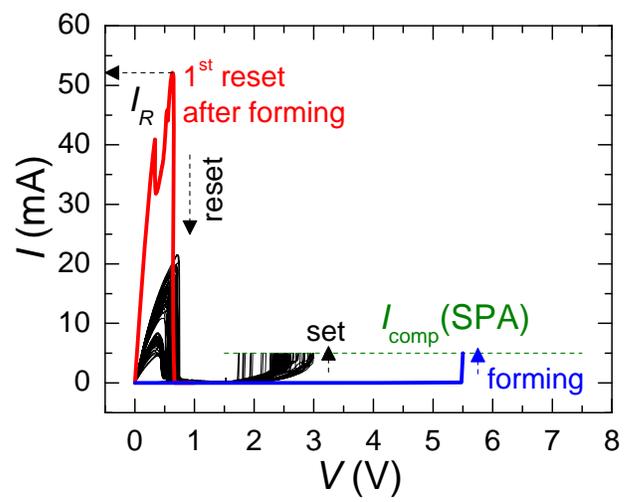

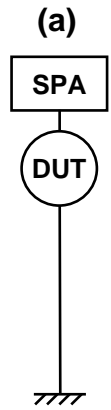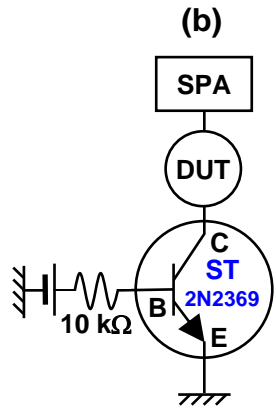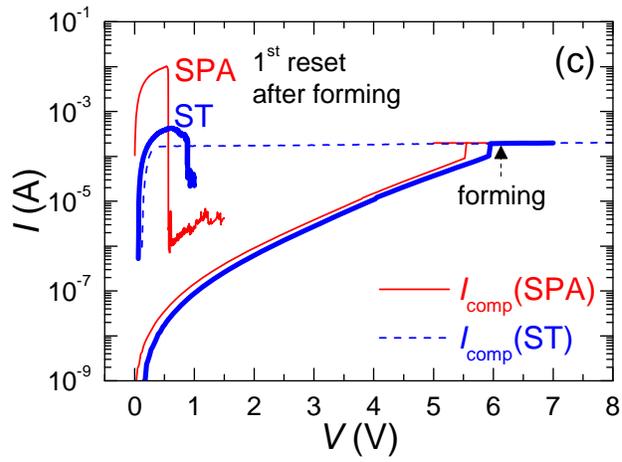

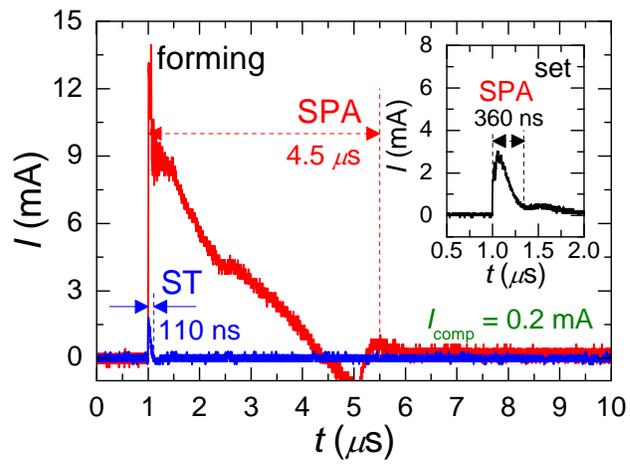

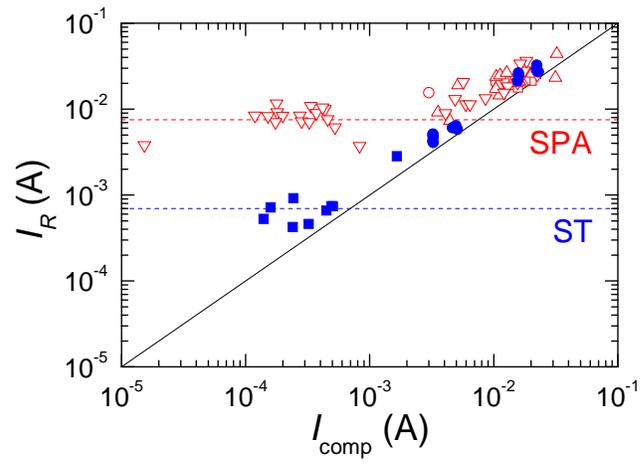